\documentclass[modern]{aastex631}
\usepackage[english]{babel}

\usepackage[letterpaper,top=2cm,bottom=2cm,left=3cm,right=3cm,marginparwidth=1.75cm]{geometry}

\usepackage{amsmath}
\usepackage{graphicx}
\usepackage[colorlinks=true]{hyperref}

\newcommand{\beq}{\begin{equation}}
\newcommand{\eeq}{\end{equation}}

\begin{document}

\title{Comment on ``Quantum-Mechanical Suppression of Gas Accretion by Primordial Black Holes''}
\author{James M.\ Cline}
\email{jcline@physics.mcgill.ca}
\affiliation{McGill University Department of Physics \& Trottier Space Institute, 3600 Rue University, Montr\'eal, QC, H3A 2T8, Canada}
\affiliation{CERN, Theoretical Physics Department, Geneva, Switzerland}

\begin{abstract}
It was recently claimed (\url{https://arxiv.org/pdf/2409.09081}) that accretion of ordinary
matter on black holes of mass $(6\times 10^{14}-4\times10^{19})\,$g would be inhibited by
quantum mechanical effects, namely the de Broglie wavelength of the electron being larger
than the Schwarzschild radius. However the conclusion is based on considering accretion of a
single atom over the age of the Universe.  There is no suppression of the accretion rate per
atom on such black holes.
\end{abstract}

%\maketitle
\section*{}

\cite{Loeb:2024gga} recently argued that primordial black holes (PBHs) in the mass range
$10^{14}-10^{19}$\,g cannot accrete ordinary matter, because their Schwarzschild radius is
too small to capture an electron fast enough, its wave function extending over a much larger
region. The nucleus on the other hand has a much smaller wavelength and can be absorbed by
a black hole in the usual classical picture.

\cite{Loeb:2024gga} estimates that the electron would take an additional $\tau =
10^{-11}$\,s to be captured, following the atomic nucleus, due the quantum mechanical
suppression.  Therefore the maximum rate of absorption of an atom by a black hole is
$m_n/\tau\sim 10^{-13}\,$g/s, where $m_n$ is the mass of the nucleus (taken to be a
proton).  This is of course a negligible rate, since it involves only a single proton. 
However the author claims on this basis that conventional hydrodynamic treatments of
accretion are invalid.  Clearly, a time delay of $10^{-11}$\,s in the process of
absorption of an atom cannot invalidate a classical description of fluid absorption, so this
argument makes no sense.

\bibliography{sample}

\end{document}